\documentclass[12pt,a4paper]{article}

\usepackage{multirow}
\usepackage{a4wide}
\usepackage{epsfig}
\usepackage{graphicx}
\usepackage{multicol}
\usepackage{slashed}
\usepackage{cancel}
\usepackage[intlimits]{amsmath}
\usepackage{amssymb}
\usepackage[small]{caption}
\usepackage{cite}
\usepackage[usenames,dvipsnames]{color}
     \definecolor{hgreen}{rgb}{0,.3,0}
     \definecolor{hred}{rgb}{.3,0,0}
     \definecolor{hblue}{rgb}{0,0,.3}
     \definecolor{LightGray}{gray}{0.95}
\usepackage[backref=page,
            colorlinks=true,
            linkcolor=hblue,
            citecolor=hgreen,
            filecolor=hblue,
            urlcolor=hred]{hyperref}

\renewcommand*{\backref}[1]{}
\numberwithin{equation}{section}

\newcommand{\xhw}{x_{hW}}
\newcommand{\xth}{x_{th}}
\newcommand{\xtz}{x_{tZ}}
\newcommand{\xwh}{x_{Wh}}
\newcommand{\xhz}{x_{hZ}}
\newcommand{\xzh}{x_{Zh}}

\providecommand{\doi}[1]{\href{http://dx.doi.org/#1}{doi:#1}}

\newcommand{\refeq}[1]{Eq.~(\ref{eq:#1})}

\allowdisplaybreaks

\begin{document}

%%%%%%%%%%%%%%%%%%%%%%%%%%%%%%%%%%%%%%%%%%%%%%%%%%%%%%%%%%%%%%%%%%%%%%%%%%

\title{ \boldmath 
\textbf{Experimental constraints on the coupling of the Higgs boson to
  electrons}\unboldmath }

\author{
{$\text{Wolfgang Altmannshofer}^{a}$, $\text{Joachim Brod}^{b}$, $\text{Martin Schmaltz}^{c}$}\\[2em]
  {\normalsize $^{a}$Perimeter Institute for Theoretical Physics, Waterloo, ON, N2L 2Y5, Canada}\\[0.5em] 
  {\normalsize $^{b}$PRISMA Cluster of Excellence and Mainz Institute
    for Theoretical Physics,}\\[0.5em] 
  {\normalsize Johannes-Gutenberg-Universit\"at, 55099 Mainz, Germany}\\[0.5em] 
  {\normalsize $^{c}$Physics Department, Boston University, Boston, MA 02215, USA}
}

\date{}

%% \begin{flushright}   
%% \begin{tabular}{l}   
%% MITP/15-010 
%% \end{tabular}   
%% \end{flushright}   

\maketitle

\thispagestyle{empty}

%%%%%%%%%%%%%%%%%%%%%%%%%%%%%%%%%%%%%%%%%%%%%%%%%%%%%%%%%%%%%%%%%%%%%%%%%%

\begin{abstract}
\addcontentsline{toc}{section}{Abstract}

In the standard model (SM), the coupling of the Higgs boson to
electrons is real and very small, proportional to the electron
mass. New physics could significantly modify both real and imaginary
parts of this coupling. We discuss experiments which are sensitive to
the Higgs-electron coupling and derive the current bounds on new
physics contributing to this coupling. The strongest constraint
follows from the ACME bound on the electron electric dipole moment
(EDM). We calculate the full analytic two-loop result for the electron
EDM and show that it bounds the imaginary part of the Higgs-electron
coupling to be less than $ 1.7 \times 10^{-2}$ times the SM electron
Yukawa coupling. Deviations of the real part are much less
constrained. We discuss bounds from Higgs decays, resonant Higgs
production at electron colliders, Higgs mediated $B\rightarrow e^+
e^-$ decays, and the anomalous magnetic moment of the
electron. Currently, the strongest constraint comes from $h \to
e^+e^-$ at the LHC, bounding the coupling to be less than $\sim 600$
times the SM Yukawa coupling. Important improvements can be expected
from future EDM measurements as well as from resonant Higgs production
at a next-generation high-luminosity $e^+ e^-$ collider.

\end{abstract}

%%%%%%%%%%%%%%%%%%%%%%%%%%%%%%%%%%%%%%%%%%%%%%%%%%%%%%%%%%%%%%%%%%%%%%%%%%

\newpage

%%%%%%%%%%%%%%%%%%%%%%%%%%%%%%%%%%%%%%%%%%%%%%%
\section{Introduction}
\label{sec:introduction}
%%%%%%%%%%%%%%%%%%%%%%%%%%%%%%%%%%%%%%%%%%%%%%%

The announcement of the five-sigma discovery of the Higgs boson at the
LHC on July 4th 2012 officially launched a new program of precision
tests of the standard model (SM) in Higgs physics. By precisely
measuring the production and decay rates of the Higgs boson we aim to
test if the Higgs' couplings agree with SM predictions, and -- if they
don't agree -- we hope to obtain hints about physics beyond the SM. 

So far, the focus has been on the largest couplings of the Higgs such
as the couplings to $W$ and $Z$ gauge bosons as well as the top,
bottom and $\tau$ Yukawa couplings~\cite{Khachatryan:2014jba,
  atlas:higgs}\footnote{Recently, methods for measuring first and
  second generation quark Yukawa couplings were proposed as
  well~\cite{Bodwin:2013gca, Delaunay:2013pja, Kagan:2014ila}.}.  In
this article, we concentrate instead on the coupling of the Higgs to
electrons which is of course predicted to be one of the smallest
couplings of the Higgs in the SM. We ask what we know about the
electron Yukawa coupling from a purely experimental point of view. It
might be reasonable to expect that new physics in the Higgs sector
couples more strongly to top quarks than to electrons; however,
measuring the Higgs coupling to electrons is interesting precisely
because the SM prediction for the Yukawa coupling is so small. A
higher-dimensional operator from new physics can easily compete with
the SM Yukawa coupling or can even dominate.

In Section~\ref{sec:coupling} we briefly discuss how
higher-dimensional operators can modify the coupling of the Higgs to
electrons. In Section~\ref{sec:direct} we analyze the sensitivity to a
modified Higgs-electron coupling coming from searches for Higgs decays
into $e^+e^-$ at the LHC and at future hadron colliders, as well as
from Higgs production at electron-positron colliders. Indirect
constraints on a modified Higgs-electron coupling from the electric
dipole moment (EDM) and the anomalous magnetic dipole moment (MDM) of
the electron, as well as from rare $B$-meson decays into $e^+e^-$
final states are discussed in Section~\ref{sec:indirect}.  We conclude
in Section~\ref{sec:conclusions} with a summary of current and future
constraints.  In Appendix~\ref{sec:twoloopEDM} we provide analytic
expressions for the complete set of relevant two-loop contributions to
the electron EDM and MDM that are induced by a modified Higgs-electron
coupling. In Appendix~\ref{sec:loophole} we show that in the Standard
Model the Higgs electron coupling is necessarily suppressed by the
electron mass to all loop orders.

%%%%%%%%%%%%%%%%%%%%%%%%%%%%%%%%%%%%%%%%%%%%%%%
\section{The Higgs-electron coupling beyond the SM}
\label{sec:coupling}
%%%%%%%%%%%%%%%%%%%%%%%%%%%%%%%%%%%%%%%%%%%%%%%

Within the SM, both the electron mass $m_e$ and the Higgs-electron
coupling $g_{eeh}$ are completely determined by the Yukawa coupling
$y_e$ of the first generation leptons to the Higgs doublet $\varphi$,
\begin{equation}
\mathcal{L}_\text{SM} \supset y_e^\text{SM} \bar{\ell}_L \varphi e_R
\, + h.c. \,.
\end{equation}
After electroweak symmetry breaking, we can parametrize $\varphi =
(G^+, (v + h + iG^0)/\sqrt{2})^T$ and thus obtain the electron mass
term and the coupling of the physical Higgs boson $h$ to left and
right handed electrons:
\begin{equation}
 \mathcal{L} \supset m_e \bar{e}_L e_R + \frac{g_{eeh}}{\sqrt{2}} \bar{e}_L e_R h \,+ h.c. \,.
\end{equation}
Given the known electron mass $m_e \simeq 0.511$ MeV and the Higgs
vacuum expectation value (vev) 
$v=(\sqrt{2} G_F)^{-1/2}\simeq 246$ GeV,
one can predict the Higgs-electron coupling in the SM
\begin{eqnarray}
\label{eq:Eyukawa}
g_{eeh}^\text{SM} = y_e^\text{SM} = \sqrt{2} m_e / v \simeq 2.9 \times 10^{-6} ~.
\end{eqnarray}
We see that the Higgs-electron coupling is of the order of $10^{-6}$
and real. At the quantum level, the magnitude of the coupling receives
well-known perturbative corrections starting at order $y_e \alpha$,
where $\alpha$ is the fine-structure constant, and requires a proper
definition of the quantities involved. We will ignore all such
complications because the experimental uncertainties will turn out to
be much larger than the change in the coupling due to running. In
Appendix~\ref{sec:loophole} we give a general proof based on chiral symmetry
showing that all quantum corrections to the Higgs-electron coupling in the SM
are proportional to the electron mass and are therefore small.

%%%%%%%%%%%%%%%%%%%%%%%%%%%%%%%%%%%%%%%%%%%%%%%
\begin{figure}[t]
\begin{center}
\includegraphics[width=0.28\textwidth]{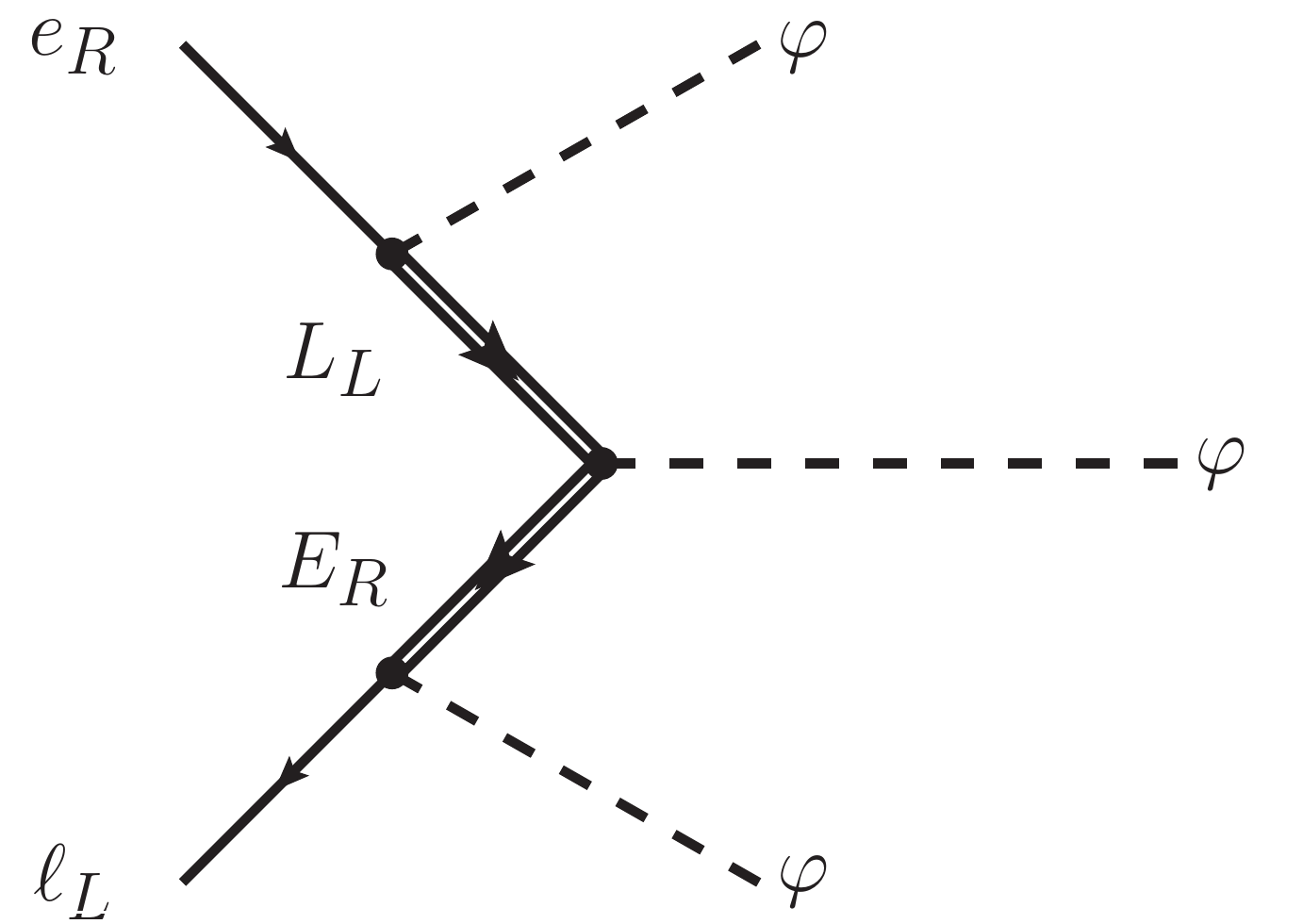}~~~~~
\includegraphics[width=0.3\textwidth]{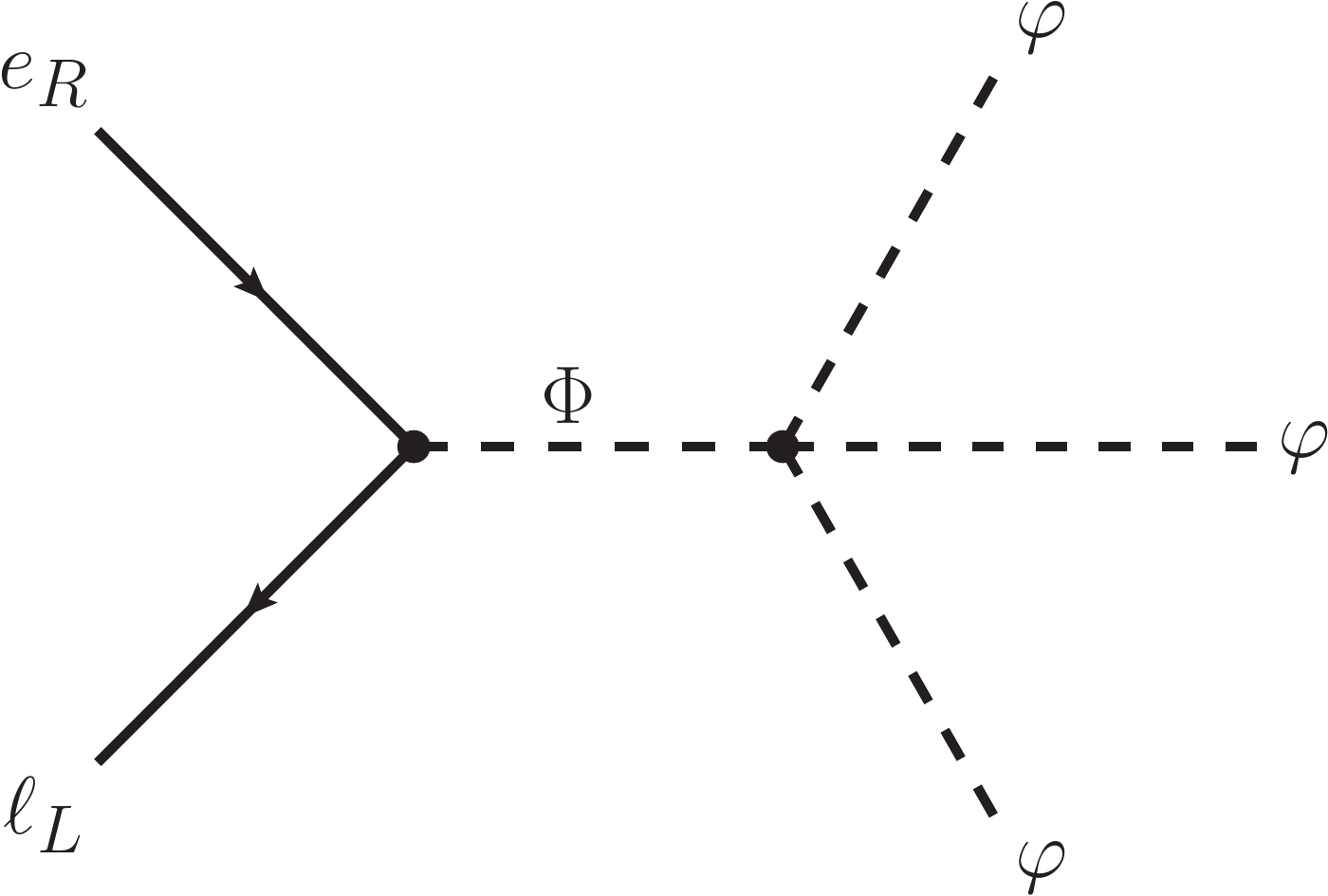}~~~~~
\includegraphics[width=0.3\textwidth]{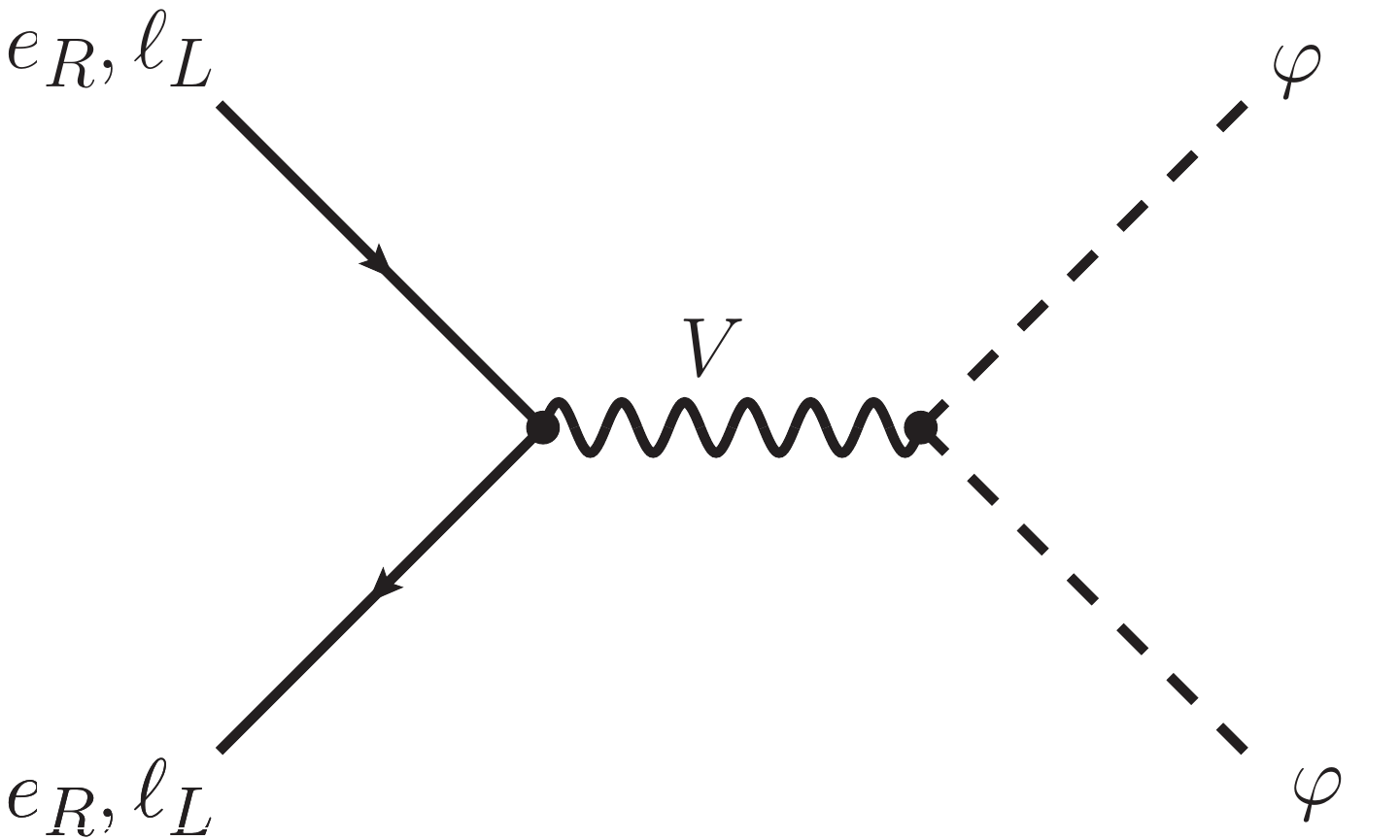}
\end{center}
\caption{Possible origins of the dimension-six operator in
  Eq.~\eqref{dim6}. Left: mixing of the electrons with heavy
  vector-like leptons. Middle: mixing of the SM Higgs doublet with a
  heavy scalar doublet that couples to electrons. Right: exchange of a
  heavy vector.
\label{fig:dim6}}
\end{figure}
%%%%%%%%%%%%%%%%%%%%%%%%%%%%%%%%%%%%%%%%%%%%%%%

How can the coupling of the Higgs to electrons differ from the value
predicted in \refeq{Eyukawa}? Assuming that the field content of the
SM provides an adequate description of physics at the weak scale, any
new physics contributions can be parametrized by higher-dimensional
operators respecting the SM gauge symmetries.  Thus, to modify the
Higgs-electron coupling, we must introduce a higher-dimensional
operator coupling the Higgs to electrons that changes the relationship
between the electron mass and the Yukawa coupling. The
lowest-dimension operators which do this are of dimension six and have
zero, one, or two derivatives
\begin{equation} \label{dim6}
\begin{split}
 \mathcal{L}_\text{dim6} &\supset \frac{c_0}{M^2} \varphi^\dagger \varphi \bar{\ell}_L \varphi e_R +h.c. \\
&+\frac{c_{1L}}{M^2} \bar{\ell}_L \gamma^\mu \ell_L \partial_\mu (\varphi^\dagger \varphi)
+\frac{c'_{1L}}{M^2} \bar{\ell}_L \gamma^\mu \ell_L\, ({\varphi^\dagger}\!\!
\stackrel{\leftrightarrow}{D}_\mu\! \varphi )
+ (\ell_L \leftrightarrow e_R) \\
&+\frac{c_2}{M^2}  \bar{\ell}_L e_R D^2  \varphi  + \ldots  
\end{split}
\end{equation}
where $M$ is a new-physics scale, $c_0$ and $c_2$ are complex
couplings, and the $c_{1}$ couplings are real. Such operators could
arise from mixing of the leptons with heavy vector-like
fermions, from mixing of the Higgs with a heavy scalar doublet, or
from the exchange of new vector bosons (see Fig.~\ref{fig:dim6}).
Generically, we would expect that the couplings $c_i$ are $3 \times 3$
matrices in lepton flavor space such that the operators in
Eq.~\eqref{dim6} not only modify the Higgs-electron coupling but also
alter the other Higgs-lepton couplings, thereby also inducing
lepton-flavor violating Higgs couplings. Possible relations among the
new physics effects in these couplings are, however, model dependent
and their discussion is beyond the scope of this work.

We now argue that the only dimension-six operator which can
significantly modify the coupling of on-shell electrons to the Higgs
is the one proportional to $c_0$.  To start, note that the operators
in the second line preserve chiral symmetry and might therefore be
expected to have a larger coefficient than those in the first and
third line which do break chiral symmetry. Yet, the operators
proportional to $c'_{1L}$ and $c'_{1R}$ do not contribute to the
Higgs-electron coupling. The operators proportional to $c_{1L}$ and
$c_{1R}$ do modify the real part of the Higgs-electron coupling;
however, after integrating by part and applying the equations of
motion one sees that these contributions are suppressed by a factor
$m_e/M$ in addition to $v/M$ and are therefore too small to be
interesting. In addition, there are several potentially interesting
two-derivative operators, but only the one shown in the third line of
Eq.~\eqref{dim6} gives contributions to the Higgs-electron coupling
which are not suppressed by powers of the electron mass. In Higgs
production or decay, the Higgs boson is on shell and the derivatives
can simply be replaced by $M_h^2$, thus allowing the effects of this
operator to be absorbed by a shift of $c_0$. For low-energy
experiments the derivatives get replaced by small momenta and the
effects of the $c_2$ operator are negligible.  We will therefore
concentrate on the $c_0$ operator as a plausible source of observable
deviations in the Higgs-electron coupling from now on.

Expanding the Higgs doublet about its vev, the $c_0$ operator in
Eq.~\eqref{dim6} leads to corrections to the electron mass and the
Higgs-electron coupling of order $v^2/M^2$:
\begin{equation}\label{mod}
\begin{split}
m_e &= \frac{v}{\sqrt{2}}\left( y_e + \frac{c_0}{2} \frac{v^2}{M^2} \right) \,, \\
g_{eeh} &=   y_e + \frac{3c_0}{2} \frac{v^2}{M^2} = \frac{\sqrt{2} m_e}{v} + c_0 \frac{v^2}{M^2} \,.
\end{split}
\end{equation}
Note the factor of 3 in the ratio of the new physics contributions to
$g_{eeh}$ and $m_e$ relative to the SM contributions. In the presence
of both the SM Yukawa coupling and the dimension-six operator, the
Higgs-electron coupling and the electron mass become independent
parameters\footnote{On the other hand, if one can neglect
  contributions of operators with mass dimension higher than six, the
  effective couplings of electrons to more than one Higgs boson are
  fixed in terms of $g_{eeh}$ and $m_e$.}. As $m_e \ll v$, the new
physics correction to the Higgs-electron coupling can be sizable, even
for very large new physics scales $M \gg v$. However, one should keep
in mind that $g_{eeh} \gg g_{eeh}^\text{SM}$ is only possible if there
is a significant cancellation between the contributions to the
electron mass coming from the Yukawa coupling and the
higher-dimensional operator, cf. Eq.~\eqref{mod}.

Note that given the smallness of the electron mass, operators of
dimension greater than 6 may also play a role in determining the
Higgs-electron coupling and the electron mass.  For instance, in the
models by Giudice and Lebedev~\cite{Giudice:2008uua} $g_{eeh}$ is
dominated by contributions from dimension-ten operators, and could be
(naturally) a factor of ${\mathcal O}(10)$ larger than the SM
prediction.

Finally we point out that $g_{eeh}$ can in general be complex; a
non-vanishing imaginary part of $g_{eeh}$ would be a clear sign of new
physics. For a sizable phase to arise there have to be at least two
different operators contributing to $g_{eeh}$, with coefficients of
similar magnitude and different phases (for instance, the
dimension-four and -six contributions in Eq.~\eqref{mod} with $y_e
\sim c_0 v^2/ M^2$). The electron mass term can always be made real by
an appropriate choice of the phase of the electron fields. The
Higgs-electron interaction then has, in general, a complex phase
relative to the mass term.

In order not to commit ourselves to a specific scenario, we find it
convenient to parametrize the modified Higgs-electron coupling more
generally as
\begin{equation} \label{eq:LagYebroken}
 g_{eeh} = \kappa_e \frac{\sqrt{2} m_e}{v} \,, 
\end{equation}
where $ \kappa_e $ is a complex parameter describing the relative
deviation from the SM prediction $\kappa_e^\text{SM}=1$. In the case
that only dimension-six operators are relevant, we can use the
relation $\kappa_e = 1 + c_0 v^3/(\sqrt{2} m_e M^2)$ together with the
assumption that $c_0$ is a coefficient of order unity to translate a bound on $\kappa_e$
into a lower bound on the NP scale $M$.

Throughout this article we set all couplings of the Higgs boson to
particles other than the electron to their SM values.

%%%%%%%%%%%%%%%%%%%%%%%%%%%%%%%%%%%%%%%%%%%%%%%
\section{Constraints from direct searches} \label{sec:direct}
%%%%%%%%%%%%%%%%%%%%%%%%%%%%%%%%%%%%%%%%%%%%%%%

The coupling of the Higgs to electrons leads to the decay of the Higgs
into electrons. Moreover, it allows resonant production of Higgs
bosons in electron-positron collisions in the $s$-channel.  In this
section we will discuss the sensitivity to a modified Higgs-electron
coupling of searches for $h \to e^+e^-$ decays at hadron colliders and
of $s$-channel Higgs production at $e^+e^-$ colliders.

%%%%%%%%%%%%%%%%%%%%%%%%%%%%%%%%%%%%%%%%%%%%%%%
\subsection{Higgs decays at the LHC and beyond}
%%%%%%%%%%%%%%%%%%%%%%%%%%%%%%%%%%%%%%%%%%%%%%%

The recent search for SM Higgs decays in the $\mu^+ \mu^-$ and $e^+
e^-$ channels by CMS~\cite{Khachatryan:2014aep} allows to set a bound
on the Higgs-electron coupling. Modifying the Higgs-electron coupling
will change both the $h \to e^+ e^-$ partial width and the total Higgs
decay width. Accordingly, we find for the modified branching ratio
\begin{equation}
\text{Br}(h \to e^+ e^-) = \frac{|\kappa_e|^2 \, \text{Br}(h \to e^+
  e^-)_\text{SM}}{1+(|\kappa_e|^2 - 1) \, \text{Br}(h \to e^+ e^-)_\text{SM}} \,,
\end{equation}
where we neglected terms that are further suppressed by $m_e^2 /
M_h^2$.  For a Higgs mass of $M_h = 125.7$~GeV~\cite{Agashe:2014kda},
the SM prediction for the branching ratio
reads~\cite{Heinemeyer:2013tqa}
\begin{equation}
 \text{Br}(h \to e^+ e^-)_\text{SM} \simeq 5.1 \times 10^{-9}\,.
\end{equation}
Assuming the SM Higgs production cross section, CMS finds an upper
bound on the branching ratio of~\cite{Khachatryan:2014aep}
\begin{equation}
 \text{Br}(h \to e^+ e^-) < 0.0019 ~~@~95\%~\text{C.L.}\,.
\end{equation}
This results in the constraint 
\begin{equation} \label{constraint}
 |\kappa_e| < 611\,.
\end{equation}
Setting the new physics coupling of the dimension-six operator
in~\eqref{dim6} to $c_0 = 1$ we can translate the constraint on
$\kappa_e$ into a constraint on the new physics scale $M$. We find $M
> 5.8$~TeV.  We expect that the bound on $\kappa_e$ from $h\to e^+e^-$
can be improved in the future at the LHC.  The gluon fusion Higgs
production cross section increases by approximately by a factor 2.5
going from 8~TeV to 14~TeV~\cite{Heinemeyer:2013tqa,
  Dittmaier:2011ti}.  Assuming that the sensitivity to the $h \to
e^+e^-$ decay scales with the square root of the number of Higgs
events, we expect sensitivities to $|\kappa_e| \sim 260 $ with 300/fb
and $|\kappa_e| \sim 150 $ with 3/ab.  At a 100 TeV proton-proton
collider the Higgs production cross section increases by another
factor of $\sim 15$~\cite{ggF}. An integrated luminosity of 3/ab might
allow to improve the sensitivity down to $|\kappa_e|\sim 75$.

We close this subsection by considering the Higgs two-body decay to
positronium and a photon. This decay is the electron analogue of the
Higgs decays to vector meson and photon which were recently
studied~\cite{Bodwin:2013gca, Delaunay:2013pja, Kagan:2014ila} to
measure the Higgs couplings to light quarks.  The idea behind this
method is that the vector meson plus gamma final state can result from
two different amplitudes which interfere. One of the amplitudes
involves the Higgs coupling to the light quarks in the vector meson,
the other amplitude generates the vector meson via mixing with a
virtual photon. Naively, these two amplitudes have different chiral
symmetry properties and cannot interfere. However, chiral symmetry is
broken dynamically by the QCD condensate, allowing the interference
term to be proportional to only one power of the small quark Yukawa
coupling.  The case of Higgs decay to positronium is quite analogous
except that here the only source of chiral symmetry breaking is the
electron mass so that the interference term is necessarily
proportional to the electron mass (times powers of alpha) in addition
to the Higgs-electron coupling. Thus this final state is less
sensitive to the Higgs-electron coupling than the $h\rightarrow e^+
e^-$ decay considered above.

%%%%%%%%%%%%%%%%%%%%%%%%%%%%%%%%%%%%%%%%%%%%%%%%%%%%%%%%%%%%%%%%%%%%%%%%%%
\subsection{Higgs production at \texorpdfstring{$e^+e^-$}{e+e-} colliders}
%%%%%%%%%%%%%%%%%%%%%%%%%%%%%%%%%%%%%%%%%%%%%%%%%%%%%%%%%%%%%%%%%%%%%%%%%%

%%%%%%%%%%%%%%%%%%%%%%%%%%%%%%%%%%%%%%%%%%%%%%%
\begin{figure}[t]
\begin{center}
\includegraphics[width=0.28\textwidth]{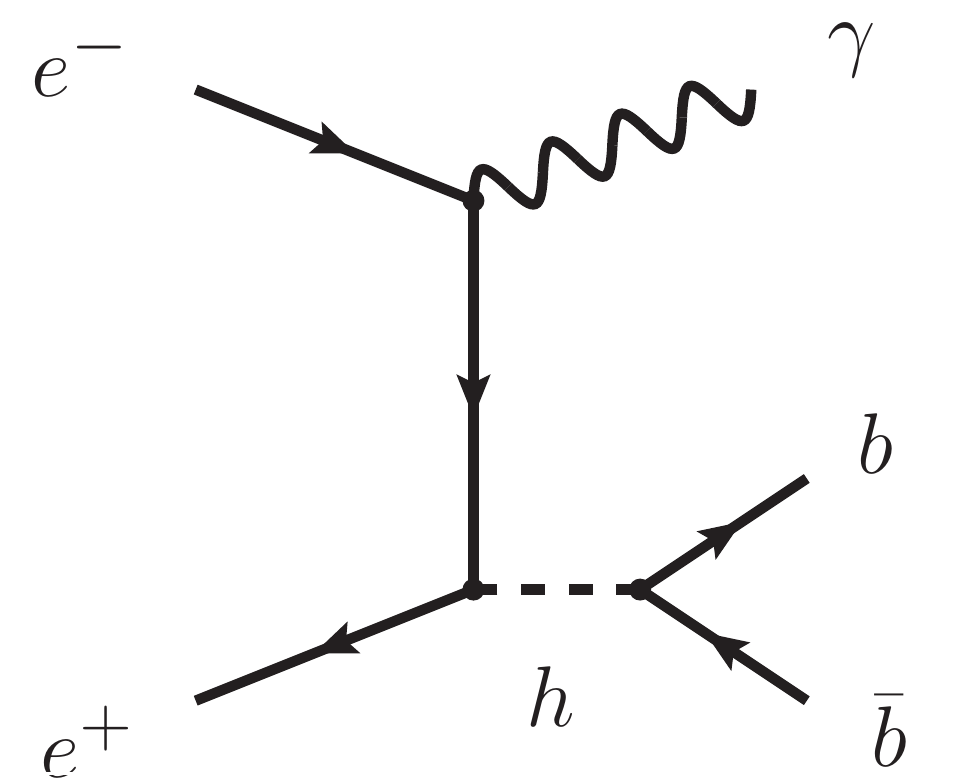}~~~~~~~~~~~~~~~~~~~~
\includegraphics[width=0.28\textwidth]{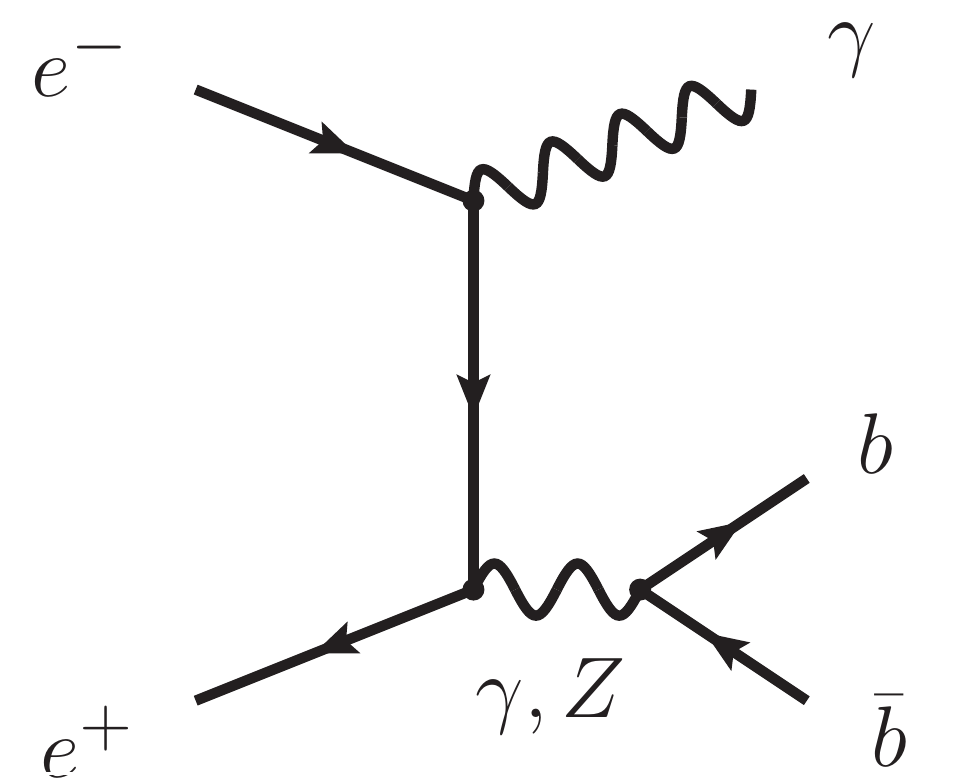}
\end{center}
\caption{ Resonant Higgs boson production at LEP~II via radiative
  return to the Higgs pole (left diagram). The Higgs is assumed to
  decay into a $b \bar b$ final state. The main background is given by
  off-shell photons or $Z$ bosons decaying into a $b \bar b$ pair
  (right diagram). \label{fig:rr}}
\end{figure}
%%%%%%%%%%%%%%%%%%%%%%%%%%%%%%%%%%%%%%%%%%%%%%%

The electron Yukawa coupling allows for resonant production of Higgs
bosons in $e^+e^-$ collisions in the $s$-channel. While the cross
section for this process is obviously maximized when the center of
mass energy is tuned to the Higgs mass, one can also obtain
sensitivity to $\kappa_e$ from virtual Higgs exchange or through
``radiative return''. Radiative return occurs when the center of mass
energy of the collider exceeds the Higgs mass; in this case
bremsstrahlung off an initial electron can reduce the effective
center-of-mass (CM) energy to the Higgs resonance.

For instance, LEP~II accumulated an integrated luminosity of the order
of 500~pb$^{-1}$ per experiment at a few different CM energies above
the Higgs pole~\cite{Alcaraz:2006mx} so that the radiative return
process was possible. To obtain a rough estimate on the reach of the
LEP~II experiments we approximate the radiative return cross section
simply as a $t$-channel process (that ignores some logarithmic
enhancement for initial-state radiation photons) with the Higgs
decaying into a $b \bar b$ pair (see Fig.~\ref{fig:rr}). We use
\texttt{madgraph}~\cite{Alwall:2014hca} to calculate the corresponding
cross section $\sigma_\text{r.r.}$, restricting the invariant mass of
the $b \bar b$ pair to the Higgs mass within the LEP~II jet energy
resolution $\sigma_{E,\text{jet}} = 10$\,GeV~\cite{Ward:1999xu}. We
further assume that the main background is provided by virtual photons
and $Z$ bosons decaying into a $b \bar b$ pair in the same
invariant-mass bin, with a cross section $\sigma_\text{bkg}$.

We collect the cross sections for various CM energies, as well as the
corresponding integrated luminosities per experiment, in
Tab.~\ref{tab:lep2}. Adding all data sets, we find $N_\text{r.r.} =
3\cdot 10^{-6} \times |\kappa_e|^2$ and $N_\text{bkg} = 121$ for the
total number of signal and background events, respectively. Setting
$N_\text{r.r.}/\sqrt{N_\text{bkg}}=1$ we see that LEP~II was, in
principle, sensitive to $|\kappa_e| \sim 2000$. We find that a similar
sensitivity could be obtained with the 20/pb that have been collected
much closer to the Higgs resonance at a center of mass energy of
130~GeV. Our rough sensitivity estimates are weaker than the LHC bound
derived in the previous section and for example do not take into account
signal efficiencies or backgrounds from fakes. The LHC bound is
expected to improve significantly after run~II. Thus, a more
sophisticated analysis of the LEP~II data does not seem worth while.

\renewcommand{\arraystretch}{1.2}
%%%%%%%%%%%%%%%%%%%%%%%%%%%%%%%%%%%%%%%%%%%%%%%%%%%%%%%%%%%%%%%
\begin{table}%[tbh] 
\begin{center}
\begin{tabular}{cccc}
\hline\hline
$E$ [GeV] & ${\cal L}$ [1/pb] & $10^6 / |\kappa_e|^2 \times \sigma_\text{r.r.}$ [fb] &
$\sigma_\text{bkg}$ [fb] \\
\hline\hline
189 & 170 & 1.40 & 56.9 \\
192 & 30  & 1.33 & 54.2 \\
196 & 80  & 1.25 & 50.8 \\
200 & 80  & 1.18 & 47.4 \\
202 & 40  & 1.14 & 45.8 \\
205 & 80  & 1.08 & 43.4 \\
207 & 140 & 1.04 & 41.9 \\
\hline\hline
\end{tabular}
\end{center}
\caption{\small The integrated luminosity ${\cal L}$ collected by each
  experiment at LEP~II at various CM energies $E$, and the
  corresponding cross sections for producing a photon, plus a $b \bar
  b$ pair with an invariant mass between 115 GeV and 135 GeV, via a
  virtual Higgs ($\sigma_\text{r.r.}$) or an off-shell photon or $Z$
  boson ($\sigma_\text{bkg}$). }
\label{tab:lep2}
\end{table}
%%%%%%%%%%%%%%%%%%%%%%%%%%%%%%%%%%%%%%%%%%%%%%%%%%%%%%%%%%%%%%%

%%%%%%%%%%%%%%%%%%%%%%%%%%%%%%%%%%%%%%%%%%%%%%%
\begin{figure}[t]
\begin{center}
\includegraphics[width=0.6\textwidth]{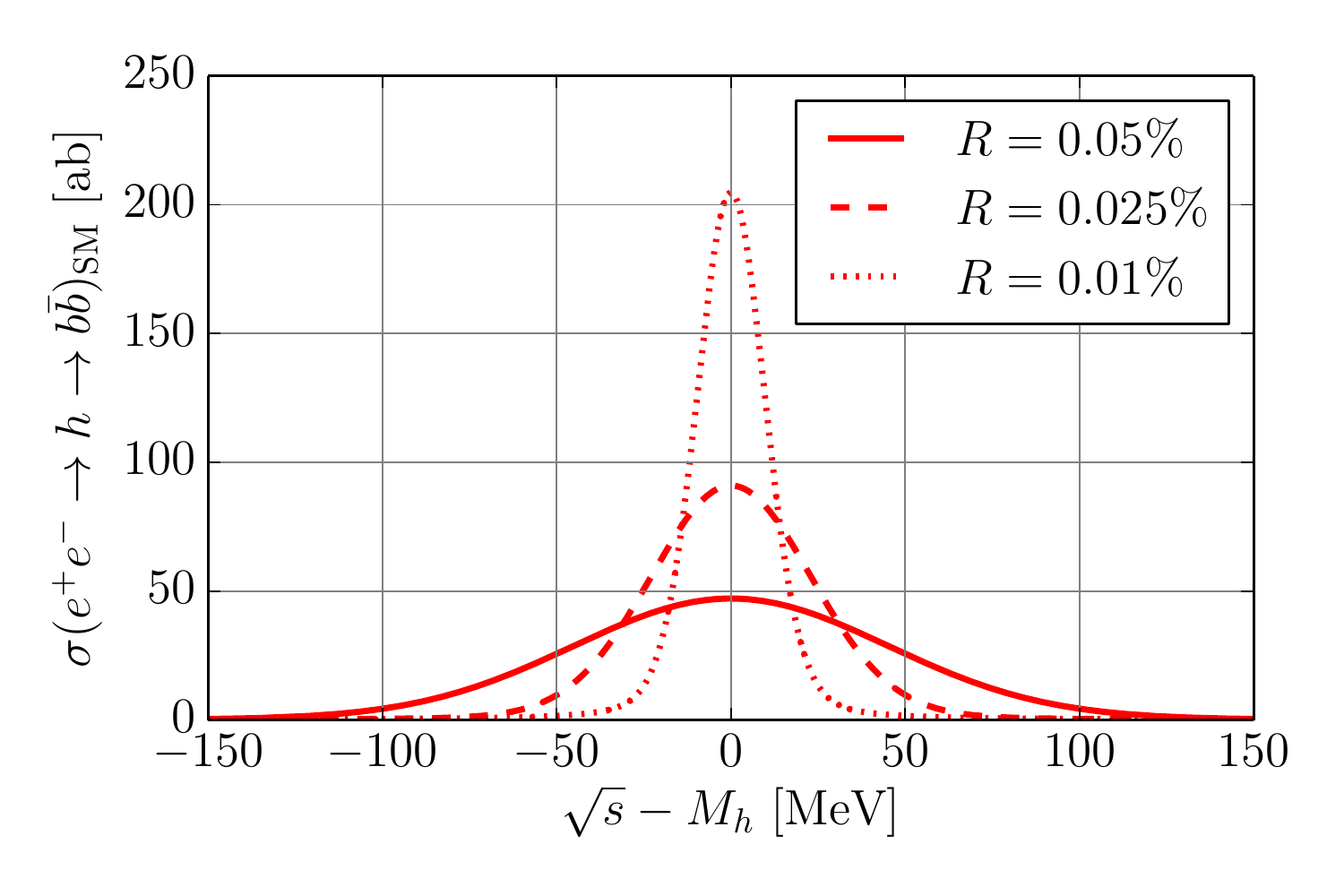}
\end{center}
\caption{Dependence of the $e^+ e^- \to h \to b \bar b$ cross section
  on the CM mass energy of the initial electron-positron
  pair. Depending on the beam energy spread $R$, the Higgs mass has to
  be known within a few to tens of MeV to fully exploit resonant
  production. \label{fig:FCCee}}
\end{figure}
%%%%%%%%%%%%%%%%%%%%%%%%%%%%%%%%%%%%%%%%%%%%%%%

On the other hand, resonant Higgs production would be possible at a
potential future $e^+ e^-$ collider running at a CM energy tuned to
the Higgs mass. The cross section for the production of a massless
fermion - antifermion pair via an $s$-channel Higgs is given by
\begin{equation}
\begin{split}
  \sigma_{e^+ e^- \to h \to f \bar f}(s) = 
  \frac{1}{32\pi} \frac{\big(y_e^\text{SM}\big)^2 y_f^2}{4} N_c^f |\kappa_e|^2 
  \frac{s}{\big(s-M_h^2\big)^2 + \Gamma_h^2 M_h^2}\,,
\end{split}
\end{equation}
where $N_c^f$ is the color factor for the final state fermions
($N_c^f=3$ for quarks and $N_c^f=1$ for leptons).  In the SM, the
width of a $125.7$~GeV Higgs is $\Gamma_h^\text{SM} =
4.17$~MeV~\cite{Heinemeyer:2013tqa}. Due to the tiny SM $h \to e^+
e^-$ branching fraction the change in the total width of the Higgs for
$\kappa_e \neq 1$ is completely negligible, given currently allowed
values of $\kappa_e$. Indeed, from the constraint in
Eq.~\eqref{constraint} we find
\begin{equation}
 \Delta \Gamma_h = \Gamma_h^\text{SM} \times (|\kappa_e|^2 - 1) \,
 \text{Br}(h \to e^+ e^-)_\text{SM} < 7.9 ~ \text{keV}\,. 
\end{equation}
In order to calculate the resonant cross section we need to convolve
the parton-level cross section $\sigma(e^+ e^- \to h \to f \bar f)$
with the beam energy resolution. We take it as a Gaussian with
variance $\Delta \equiv R \sqrt{s} / \sqrt{2}$, where $R$ is the
percentage beam energy resolution~\cite{Han:2012rb}.  Using
$R=0.05\%$~\cite{Gomez-Ceballos:2013zzn} and assuming an average
center of mass energy exactly at the Higgs mass we find the following
signal cross section for bottom quarks in the final state
\begin{equation}
\sigma_\text{sig}(e^+e^- \to h \to b\bar b) \simeq 
|\kappa_e|^2  \times  0.05/\text{fb} \,.
\end{equation}
For 100/fb of data at the Higgs resonance, this corresponds to
approximately $N_\text{sig} \simeq 5 \times |\kappa_e|^2$ signal
events.

The main background will be $f \bar f$ production via an intermediate
photon or $Z$ boson. The corresponding total cross section
is~\cite{Consoli:1989pc}
\begin{equation}
\begin{split}
  \sigma_{e^+ e^- \to \gamma, Z \to f \bar f}(s) =
  \frac{4\pi\alpha^2}{3s} N_c^f \bigg[ Q_f^2 +
    \frac{(v_e^2+a_e^2)(v_f^2+a_f^2)s^2 - 2v_e v_f Q_f s (s-M_Z^2)}{
      \big(s-M_Z^2\big)^2 + \Gamma_Z^2 M_Z^2}\bigg] \,.
\end{split}
\end{equation}
The parameters $v_f$ and $a_f$ are the vector and axial-vector
couplings of the $Z$ boson to a fermion $f$.  They are given by
\begin{equation}
  v_f = \frac{I_3^f - 2Q_f \sin^2\theta_w}{2\sin\theta_w\cos\theta_w}
  \,, \quad a_f = \frac{I_3^f}{2\sin\theta_w\cos\theta_w} \,,
\end{equation}
where $I_3^f$ and $Q_f$ denote the third isospin component and the
electric charge of the fermion~$f$, respectively. Assuming again
100/fb of data and $\sqrt{s} = M_h$ we expect roughly $N_\text{bkg} =
10^6$ $b \bar b$ background events. Requiring
$N_\text{sig}/\sqrt{N_\text{bkg}} = 1$ we estimate that one can reach
sensitivity to $|\kappa_e| \lesssim 15$ for 100/fb and to $|\kappa_e|
\lesssim 50$ for 1/fb. Slightly better sensitivities could be achieved
with a smaller beam energy spread.

Note that, in order to exploit the full benefit of resonant Higgs
production, the Higgs mass has to be known with high
precision. Fig.~\ref{fig:FCCee} shows the $e^+ e^- \to h \to b \bar b$
cross section as a function of the center of mass energy of the
initial state electrons for three choices of the beam energy
resolution $R=0.05\%$, $R=0.025\%$, and $R=0.01\%$. The cross-section
drops quickly if the center of mass energy differs from the Higgs mass
by more than a few to tens of MeV, depending on the beam energy
spread.

%%%%%%%%%%%%%%%%%%%%%%%%%%%%%%%%%%%%%%%%%%%%%%%
\section{Precision constraints} \label{sec:indirect}
%%%%%%%%%%%%%%%%%%%%%%%%%%%%%%%%%%%%%%%%%%%%%%%

We have seen that the LHC sensitivity to the Higgs electron coupling
is unlikely to reach values better than $|\kappa_e| \simeq 100$
whereas a future $e^+ e^-$ collider running on the Higgs resonance
could be sensitive to $|\kappa_e|$ of order 10.  In addition to these
direct searches, low energy precision observables can be used to
indirectly probe modified Higgs couplings.  Constraints from
low-energy flavor observables on flavor-violating fermion-Higgs
couplings have been derived for example in~\cite{Blankenburg:2012ex,
  Harnik:2012pb, Gorbahn:2014sha}. Constraints from EDMs on CP
violating top-Higgs and photon-Higgs couplings are discussed for
example in~\cite{McKeen:2012av, Brod:2013cka, Altmannshofer:2013zba}.

In this section we investigate indirect constraints on a modified
Higgs-electron coupling.  We will see that the strongest constraints
arise from the electric and magnetic dipole moments of the electron,
whereas rare $B$ decays into $e^+ e^-$ final states do not yield
competitive bounds. Note that the indirect constraints derived in this
section hold barring accidental cancellations with additional
contributions to the low energy observables that might arise in
explicit models that give rise to the higher-dimensional operators
modifying the Higgs couplings. Here we assume that all couplings other
than the Higgs-electron coupling are SM-like.
  
%%%%%%%%%%%%%%%%%%%%%%%%%%%%%%%%%%%%%%%%%%%%%%%
\subsection{Electric dipole moment of the electron} \label{sec:EDM}
%%%%%%%%%%%%%%%%%%%%%%%%%%%%%%%%%%%%%%%%%%%%%%%

%%%%%%%%%%%%%%%%%%%%%%%%%%%%%%%%%%%%%%%%%%%%%%%
\begin{figure}[t]
\begin{center}
\includegraphics[width=0.3\textwidth]{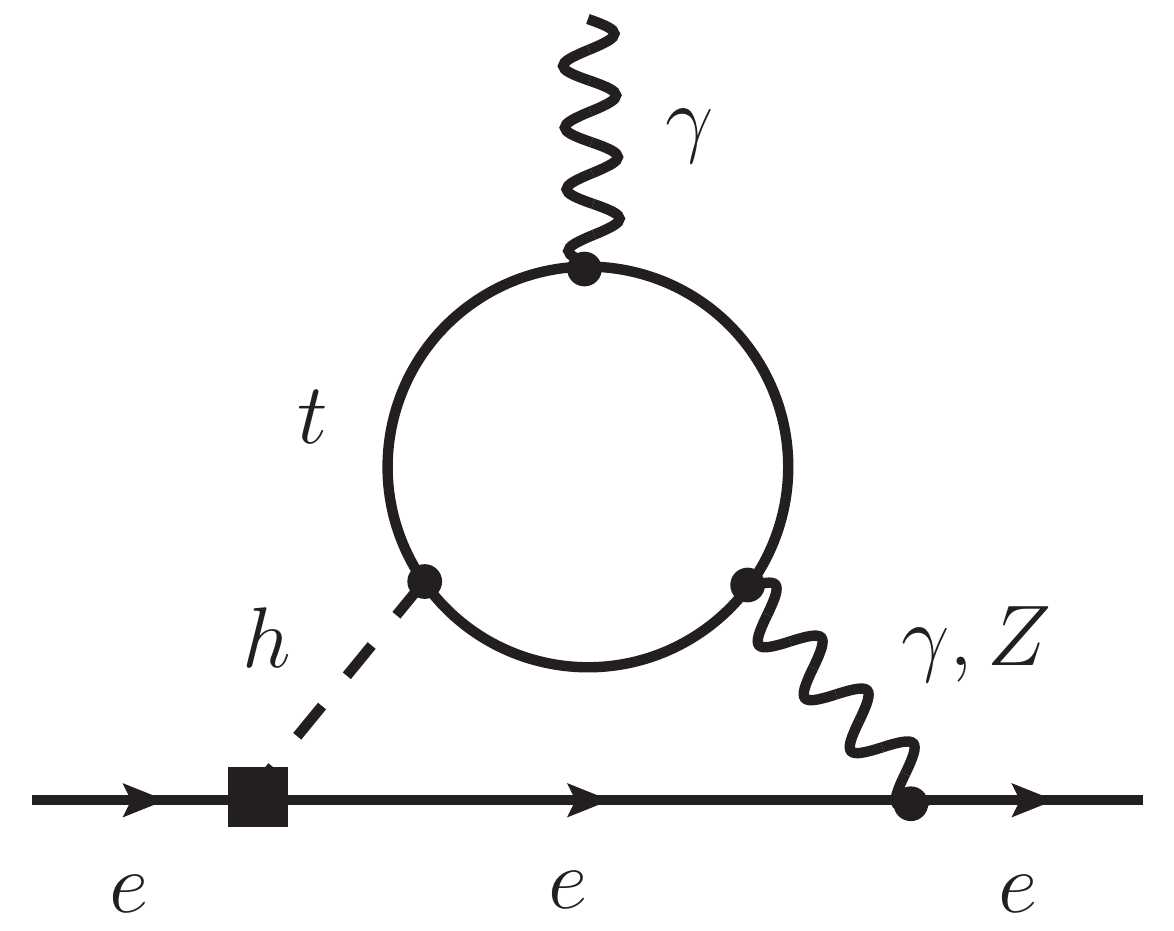}~~~~~
\includegraphics[width=0.3\textwidth]{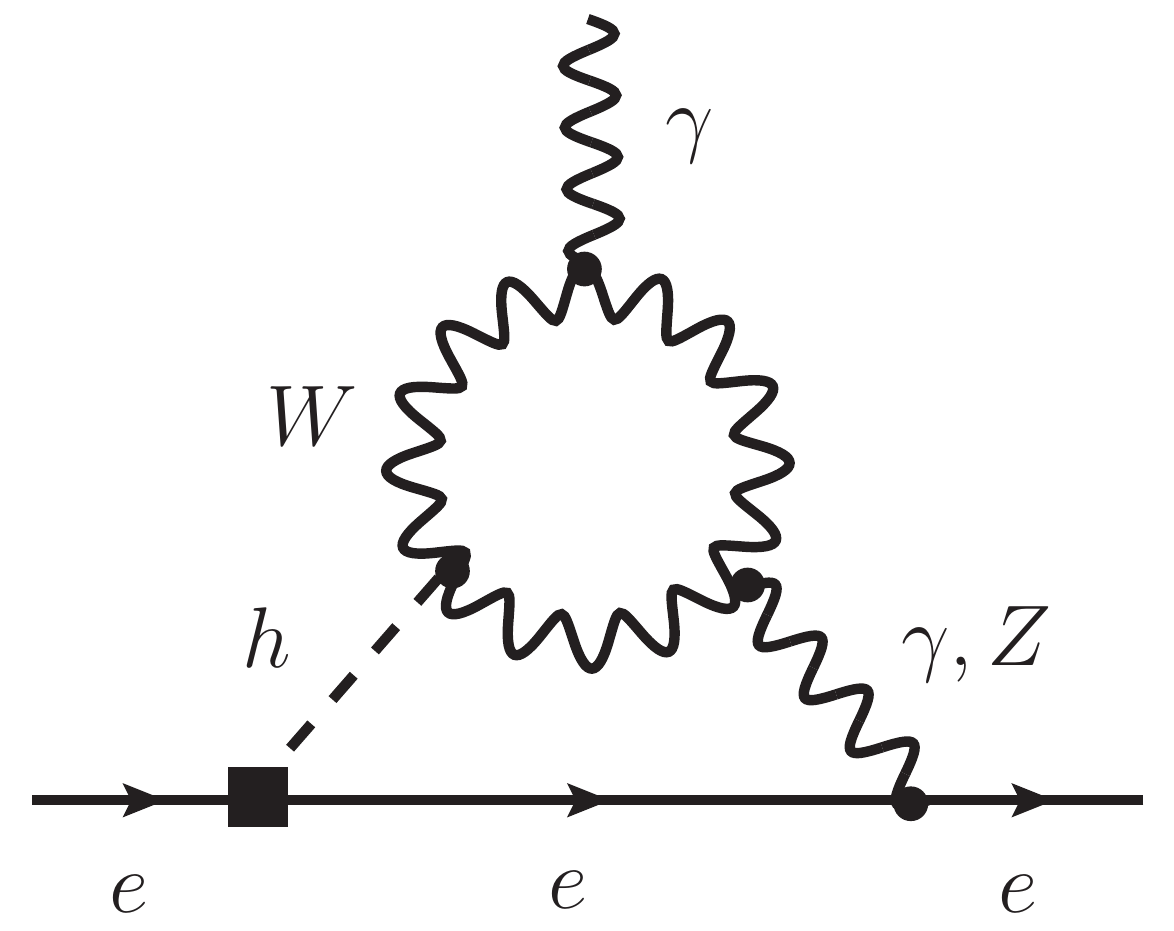}~~~~~
\includegraphics[width=0.3\textwidth]{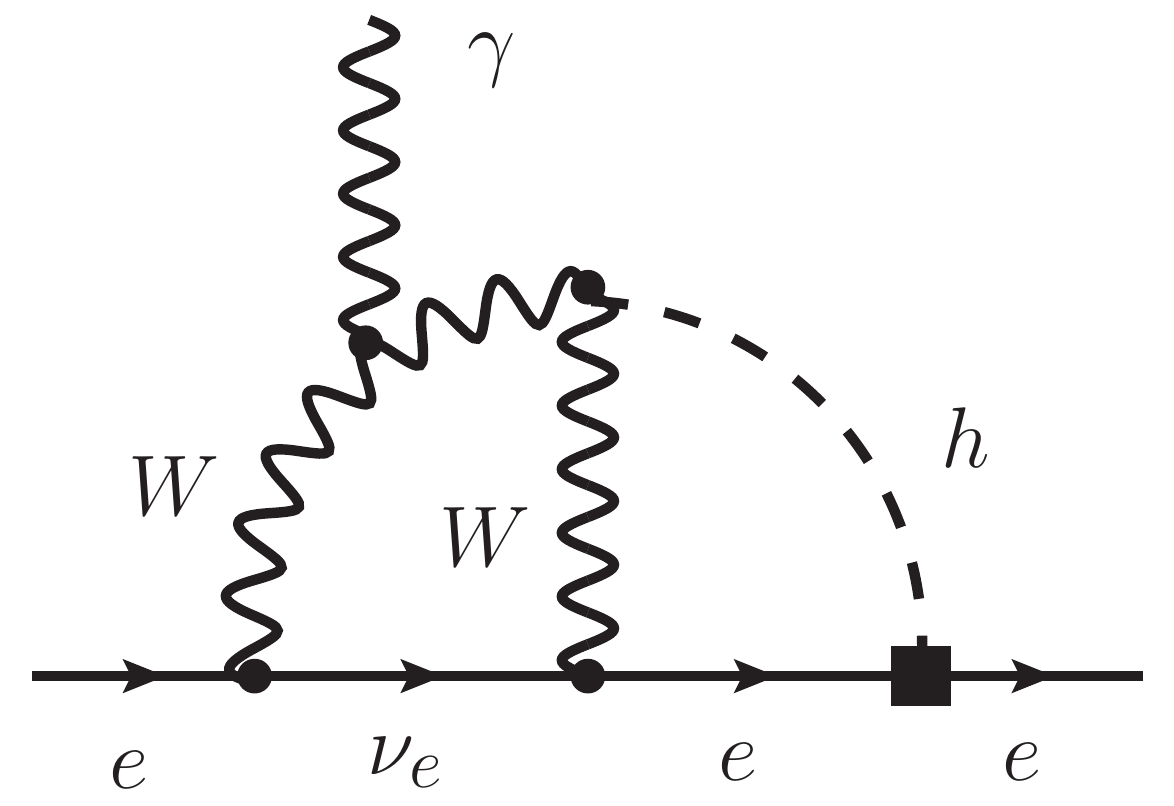}
\end{center}
\caption{Sample two-loop Feynman diagrams inducing an EDM for the
  electron through a CP-violation Higgs coupling to the electron (here
  denoted by the black square). \label{fig:2loopEDM}}
\end{figure}
%%%%%%%%%%%%%%%%%%%%%%%%%%%%%%%%%%%%%%%%%%%%%%%

The imaginary part of the Higgs boson coupling to electrons in
Eq.~\eqref{eq:LagYebroken} induces an EDM of the electron\footnote{We
  define $\sigma_{\mu\nu} = i[\gamma_\mu, \gamma_\nu]/2$. }
\begin{equation}\label{eq:edmlag}
  {\mathcal L}_\text{eff}^e = - \frac{d_e}{2} \, \bar \psi_e \,
  \sigma_{\mu\nu} \, i \gamma_5 \, \psi_e \, F^{\mu\nu}
\end{equation}
via two-loop electroweak diagrams\footnote{One-loop contributions are
  suppressed by additional powers of the electron Yukawa and electron
  mass, and are therefore negligibly small~\cite{Barr:1990vd}.} (see
Fig.~\ref{fig:2loopEDM} for sample Feynman diagrams).
We have calculated the full set of relevant two-loop contributions
that contain exactly one power of the Higgs-electrons coupling. The
analytic expressions can be found in App.~\ref{sec:twoloopEDM}. Taking
the numerical values of the input parameters ($\alpha$, $M_W$, $M_Z$,
$M_h$, $m_t$) from Ref.~\cite{Agashe:2014kda} we obtain
\begin{equation}
  \left|\frac{d_e}{e}\right| \simeq 5.1 \times |\text{Im} \kappa_e|
  \times 10^{-27} \text{cm} \,. 
\end{equation}
Using the most recent bound on the electron EDM obtained by the ACME
collaboration~\cite{Baron:2013eja}, 
\begin{equation}
 \left|\frac{d_e}{e}\right|_\text{exp} < 8.7 \times 10^{-29}
 ~\text{cm} ~~@ ~90\%~\text{C.L.} \,, 
\end{equation}
we find the very stringent constraint
\begin{equation}
  |\text{Im}\,\kappa_e| < 1.7 \times 10^{-2} ~.
\end{equation}
If the new physics contribution to the Higgs-electron coupling
contains an O(1) phase, this bound translates into a very strong
constraint on the new physics scale $M$. Setting $c_0 = i$ (in the
basis where the electron mass term is real) we find $M \gtrsim
1000$~TeV.  It is expected that the experimental sensitivity to the
electron EDM can be improved by up to two orders of magnitude in the
future~\cite{Hewett:2012ns}. Such sensitivities would allow to probe
$|\text{Im}\,\kappa_e|$ at the level of $10^{-4}$ and new physics
scales as high as $10^4$~TeV.

%%%%%%%%%%%%%%%%%%%%%%%%%%%%%%%%%%%%%%%%%%%%%%%
\subsection{Anomalous magnetic dipole moment of the electron}
%%%%%%%%%%%%%%%%%%%%%%%%%%%%%%%%%%%%%%%%%%%%%%%

The real part of $\kappa_e$ modifies the SM contribution to the
anomalous magnetic dipole moment of the electron, 
\begin{equation}
  {\mathcal L}_\text{eff}^m = - \frac{e}{4} \frac{a_e}{m_e} \, \bar \psi_e \,
  \sigma_{\mu\nu} \, \psi_e \, F^{\mu\nu} \,,
\end{equation}
via the same two-loop diagrams that induce also an EDM (see
Fig.~\ref{fig:2loopEDM} and App.~\ref{sec:twoloopEDM}). Denoting the
contributions of the two-loop diagrams with an anomalous Higgs
coupling by $\Delta a_e$, we find
\begin{equation}
 |\Delta a_e| \simeq 2.6 \times (\text{Re}\, \kappa_e - 1) \times 10^{-16} \,. 
\end{equation}
The anomaly in the gyromagnetic ratio of the electron, $a_e \equiv
(g-2)_e/2$, is conventionally used to determine the fine-structure
constant $\alpha$~\cite{Hanneke:2008tm,Aoyama:2014sxa}. However, as
pointed out in Ref.~\cite{Giudice:2012ms}, the recent precise
independent measurements of the fine-structure constant in atomic
physics experiments can be used to obtain a SM prediction for $a_e$
with an uncertainty that is only a factor of few larger than the
experimental measurement. Therefore, the anomalous magnetic moment of
the electron can be used as a probe of new physics.

We employ the value $\alpha^{-1} = 137.035999037(91)$ from the most
recent determination of the fine-structure constant using a
measurement of the ratio between the Planck constant and the mass of
the $^{87}$Rb atom~\cite{Bouchendira:2010es}.  Using the corresponding
uncertainty induced on $a_e$ around the SM value, we obtain the
allowed range for the new physics contribution to $a_e$
\begin{equation}
 |\Delta a_e| < 8.1 \times 10^{-13} \,.
\end{equation}
This translates
into the allowed range for $\kappa_e$,
\begin{equation}
   |\text{Re}\,\kappa_e| < 3.1 \times 10^{3} \,.
\end{equation}
This is a factor of five above the direct bound derived from the CMS
search for $h \to e^+e^-$. Note, however, that this bound scales
linearly with $\text{Re}\,\kappa_e$, in contrast to the quadratic
dependence of the collider constraints. The bound from the anomalous
magnetic moment can be improved in the near future by an order of
magnitude~\cite{Giudice:2012ms}, making it competitive to the expected
sensitivities from $h \to e^+e^-$ at run II of the LHC.

%%%%%%%%%%%%%%%%%%%%%%%%%%%%%%%%%%%%%%%%%%%%%%%
\subsection{Rare \texorpdfstring{$B$}{B} decays}
%%%%%%%%%%%%%%%%%%%%%%%%%%%%%%%%%%%%%%%%%%%%%%%

In the Standard Model, the rare decays $B_q \to \ell^+ \ell^-$ are
mediated by $Z$-penguin and box diagrams and require a helicity flip
of the final state leptons due to the pseudo-scalar nature of the
$B_q$ meson. Therefore, the branching ratios are proportional to the
lepton mass squared and extremely small. Higgs mediated contributions
to these decays do not, in general, suffer from the strong helicity
suppression. However, in the SM they are suppressed by the tiny lepton
Yukawa couplings and are negligible. One might therefore hope that
experiments searching for the $B_q \to e^+ e^-$ decays are sensitive
to an enhanced Higgs-electron coupling. Here we show that the current
and expected sensitivities are not competitive with the direct and
indirect bounds discussed so far.

The SM predictions for the time integrated $B_q \to e^+ e^-$ branching
ratios read~\cite{Bobeth:2013uxa}
\begin{equation}
 \text{Br}(B_s \to e^+e^-)_\text{SM} = (8.54\pm0.55)\times 10^{-14} ~,
\end{equation}
\begin{equation}
 \text{Br}(B_d \to e^+e^-)_\text{SM} = (2.48\pm0.21)\times 10^{-15} ~.
\end{equation}
These values are many orders of magnitude below the current
experimental constraints set by CDF~\cite{Aaltonen:2009vr} at 95\% C.L.
\begin{equation}
 \text{Br}(B_s \to e^+e^-) < 2.8 \times 10^{-7} ~,
\end{equation}
\begin{equation}
 \text{Br}(B_d \to e^+e^-) < 8.3 \times 10^{-8} ~.
\end{equation}
While the experimental constraints are likely to be improved at LHCb
and Belle~II by one or two orders of magnitude, sensitivities to the
SM predictions will not be reached within the foreseeable future.

In the presence of an enhanced Higgs-electron coupling, we find for
the Higgs-mediated correction to the branching ratios\footnote{Here we
  assume that $\kappa_e$ does not contain a CP violating phase. As
  discussed in section~\ref{sec:EDM}, such a phase is strongly
  constrained by the electron EDM.}
\begin{equation}
 \frac{\text{Br}(B_q \to e^+e^-)}{\text{Br}(B_q \to e^+e^-)_\text{SM}}
 - 1 ~\propto~ \frac{m_{B_q}^4}{M_h^4} \kappa_e^2 \,, 
\end{equation}
with a proportionality factor that is parametrically of order 1.  This
implies that significant enhancements of the branching ratios are only
possible for $\kappa_e \gg M_h^2/m_{B_q}^2 \sim 550$. The current
experimental constraints on $B_q \to e^+e^-$ probe couplings of the
order of $\kappa_e \sim O(10^6)$ that are already excluded by orders
of magnitude by the LHC results on $h \to e^+e^-$.

%%%%%%%%%%%%%%%%%%%%%%%%%%%%%%%%%%%%%%%%%%%%%%%%%%%%%%%%%%%%
\section{Discussion and Conclusions} \label{sec:conclusions}
%%%%%%%%%%%%%%%%%%%%%%%%%%%%%%%%%%%%%%%%%%%%%%%%%%%%%%%%%%%%

The question ``what do we know about the electron Yukawa'' is both
interesting and non-trivial to answer.

NP effects could lead to significant changes to the Higgs coupling to
electrons precisely because it is predicted to be tiny in the SM.
Enhancements of the coupling by orders of magnitude above the SM value
are theoretically possible, however only at the cost of significant
fine tuning of the electron mass. Order one changes to both the
real and imaginary parts of the coupling could be completely natural.

As a side effect, direct verification of an enhanced coupling of the
Higgs to electrons would also lead to stronger indirect constraints on
CP violating couplings of the Higgs boson to top
quarks~\cite{Brod:2013cka}.

In this article, we considered which experiments currently provide the
most stringent bounds on anomalous Higgs-electron couplings. We find
that the strongest bound on the magnitude of the coupling comes from a
CMS search for the $h \to e^+ e^-$ decay. The CP-violating imaginary
part of the Higgs-electron coupling is strongly constrained by the
current upper bound on the electron EDM. The indirect constraint on
the CP-conserving real part of the coupling from the electron $g-2$,
on the other hand, is currently relatively weak; it can, however, be
improved by a new generation of precision experiments and could be
competitive with the bounds derived from future LHC data. Finally, we
showed that rare $B$ decays are not competitive in setting bounds on
deviations from the SM Higgs-electron coupling.

Potentially the best future bounds on the magnitude of the coupling
could be obtained from an electron-positron collider running on the
Higgs resonance. With optimistic assumptions a measurement of the
Higgs-electron coupling only an order of magnitude above its SM value
seems possible.  Sensitivity to the SM value itself would require huge
amounts of statistics collected at the Higgs resonance, very precise
knowledge of the Higgs mass of the order of the Higgs width, and
exquisite control of the beam energy at the same level. It does not
seem that precision measurements of the magnitude of the SM electron
Yukawa coupling will ever be possible.

We summarize the current constraints and future expected sensitivities to a
modified Higgs-electron coupling $\kappa_e$ and the corresponding new
physics scale $M$ in Table~\ref{tab:summary}.

\renewcommand{\arraystretch}{1.6}
\setlength\tabcolsep{12pt}
%%%%%%%%%%%%%%%%%%%%%%%%%%%%%%%%%%%%%%%%%%%%%%%%%%%%%%%%%%%%%%%
\begin{table}%[tbh] 
\begin{center}
\begin{tabular}{clll}
\hline\hline
\multirow{4}{*}{$h \to e^+e^-$} & LHC8 (25/fb) & $|\kappa_e| \lesssim 600$ & $M \gtrsim 6$~TeV \\
 & LHC14 (300/fb) & $|\kappa_e| \sim 260$ & $M \sim 9$~TeV \\
 & LHC14 (3/ab) & $|\kappa_e| \sim 150$ & $M \sim 12$~TeV \\
 & 100 TeV (3/ab) & $|\kappa_e| \sim 75$ & $M \sim 17$~TeV \\
\hline
\multirow{3}{*}{$e^+e^- \to h$} & LEP~II & $|\kappa_e| \lesssim 2000$ & $M \gtrsim 3$~TeV\\
 & TLEP (1/fb) & $|\kappa_e| \sim 50$ & $M \sim 20$~TeV \\
 & TLEP (100/fb) & $|\kappa_e| \sim 10$ & $M \sim 50$~TeV \\
 \hline
\multirow{2}{*}{$d_e$} & current & Im\,$\kappa_e \lesssim 0.017$ & $M \gtrsim 1000$~TeV \\
 & future & Im\,$\kappa_e \sim 0.0001$ & $M \sim 10^4$~TeV \\
\multirow{2}{*}{$(g-2)_e$} & current & Re\,$\kappa_e \lesssim 3000 $ & $M \gtrsim 2.5$~TeV \\
 & future & Re\,$\kappa_e \sim 300 $ & $M \sim 8$~TeV \\
\hline\hline
\end{tabular}
\end{center}
\caption{\small Summary of current constraints and future expected
  sensitivities to a modified Higgs-electron coupling $\kappa_e$ and
  the corresponding new physics scale $M$.  }
\label{tab:summary}
\end{table}
%%%%%%%%%%%%%%%%%%%%%%%%%%%%%%%%%%%%%%%%%%%%%%%%%%%%%%%%%%%%%%%

%%%%%%%%%%%%%%%%%%%%%%%%%%%%%%%%%%%%%%%%%%%%%%%
\phantomsection
\addcontentsline{toc}{section}{Acknowledgments}
\section*{Acknowledgments}
%%%%%%%%%%%%%%%%%%%%%%%%%%%%%%%%%%%%%%%%%%%%%%%

This work was initiated at the Aspen Center for Physics with partial
support from the National Science Foundation, Grant No. PHYS-1066293.
J.B. acknowledges insightful discussions with Felix Yu, and support by
the U.S. National Science Foundation under CAREER Grant PHY-1151392,
the ERC Advanced Grant EFT4LHC of the European Research Council, and
the Cluster of Excellence Precision Physics, Fundamental Interactions
and Structure of Matter (PRISMA-EXC 1098). M.S. would like to thank
Andy Cohen for helpful discussions and the US Department of Energy
Office of Science for support under Award DE-SC-0010025.  Research at
Perimeter Institute is supported by the Government of Canada through
Industry Canada and by the Province of Ontario through the Ministry of
Economic Development \& Innovation.

%%%%%%%%%%%%%%%%%%%%%%%%%%%%%%%%%%%%%%%%%%%%%%%%%%%%%%%%%%%%%%%%%%%%%%%%%%

\begin{appendix}
%%%%%%%%%%%%%%%%%%%%%%%%%%%%%%%%%%%%%%%%%%%%%%%%
\section{Two-loop contributions to dipole moments}\label{sec:twoloopEDM}
%%%%%%%%%%%%%%%%%%%%%%%%%%%%%%%%%%%%%%%%%%%%%%%%

In this appendix we give the analytic expressions for the complete set
of relevant two-loop contributions to the electron dipole moments that
are induced by a modified Higgs-electron coupling. For the case
$\kappa_e = 1$ we reproduce exactly the part of the bosonic
contributions in Ref.~\cite{Gribouk:2005ee} that involve the exchange
of a virtual Higgs boson. To our knowledge, this constitutes the first
independent (partial) check of their calculation. For an imaginary
value of $\kappa_e $ our results for the top-loop diagrams with an
internal photon are in agreement with the classic calculation by Barr
and Zee~\cite{Barr:1990vd}, while the corresponding analytic results
with an internal $Z$ boson are new. Results for the considered bosonic
diagrams, in terms of parametric integrals, can in principle be
extracted from Ref.~\cite{Leigh:1990kf, Chang:1990sf} that give
results for two-loop contributions to EDMs in multi-Higgs doublet
models (see also~\cite{Abe:2013qla} for a recent reevaluation of the
Barr-Zee type contributions in two-Higgs doublet models). We find
small numerical discrepancies with the results of~\cite{Leigh:1990kf,
  Chang:1990sf} of the order of 10\%.

To obtain our results we performed an off-shell matching calculation, 
along the lines of Ref.~\cite{Bobeth:1999mk}, to an effective theory
where all heavy particles (the top quark and the $W$, $Z$, and Higgs
bosons) are integrated out. The two physical operators, yielding the
magnetic and electric dipole moments in the non-relativistic limit,
can be chosen as
\begin{equation}
 {\cal O}_m = e \bar \psi_e \sigma^{\mu\nu} \psi_e F_{\mu\nu} \,, \qquad
 {\cal O}_e = e \bar \psi_e \sigma^{\mu\nu} i \gamma_5 \psi_e F_{\mu\nu} \,.
\end{equation}
In order to project on the physical matrix elements, we also need the
following two operators that vanish via the electron equations of
motion: 
\begin{equation}
 {\cal O}_{m}^\text{e.o.m.} = \bar \psi_e \slashed{D} \slashed{D} \psi_e \,, \qquad
 {\cal O}_{e}^\text{e.o.m.} = \bar \psi_e \slashed{D} \slashed{D} i \gamma_5 \psi_e \,.
\end{equation}
In our calculation we set the electron mass to zero while keeping the
electron Yukawa nonzero. Therefore, no other off-shell operators can
contribute at this order, and it is sufficient to expand the
integrands to first order in the external momenta.

We have calculated all Feynman diagrams employing the background field
gauge for the electroweak interactions~\cite{Denner:1994xt}. The
two-loop integrals were computed using the recursion relations
in~\cite{Davydychev:1992mt, Bobeth:1999mk}. We decompose our result
for the two-loop electron EDM in the following way
(cf. Eq.~\eqref{eq:edmlag})
\begin{equation} \label{EDM}
 d_e^\text{2loop} = d_e^{t\gamma} + d_e^{tZ} + d_e^{W\gamma} + d_e^{WZ} + d_e^{W} + d_e^{Z} ~. 
\end{equation}
The first four terms denote contributions from Barr-Zee type
diagrams~\cite{Barr:1990vd} containing top-quark loops and a photon
($d_e^{t\gamma}$), top-quark loops and a $Z$ boson ($d_e^{tZ}$),
$W$ boson loops and a photon ($d_e^{W\gamma}$), and $W$ boson loops
and a $Z$ boson ($d_e^{WZ}$) (see the left and center diagrams in
Fig.~\ref{fig:2loopEDM} for examples). The last two terms in
\eqref{EDM} denote the remaining two-loop contributions that contain
either $W$ bosons ($d_e^W$), or $Z$ bosons ($d_e^Z$) (see the right
diagram in Fig.~\ref{fig:2loopEDM} for an example). We obtain for the
individual contributions
\begin{eqnarray}
 \frac{d_e^{t\gamma}}{e} &=& \frac{16 e^2}{3 (16\pi^2)^2}
 \frac{y_e^\text{SM}}{\sqrt{2} v} ~\text{Im}\,\kappa_e \xth \left[
   \left(2\xth - 1\right) \Phi\left(\frac{1}{4\xth}\right) - 2\left( 2
   + \log \xth \right)
   \right] \,, \\[16pt]
 \frac{d_e^{tZ}}{e} &=& \frac{e^2}{(16\pi^2)^2s_w^2} \frac{y_e^\text{SM}}{\sqrt{2}v}
 ~\text{Im}\,\kappa_e~ \frac{1}{2c_w^2} \left( 1 - 4 s_w^2\right)
 \left( 1 - \frac{8}{3} s_w^2\right) \left(1 - \xhz \right)^{-1} \xtz \nonumber \\[2mm]
 && \times \bigg[\left(1 - 2\xth \right)
   \Phi\left(\frac{1}{4\xth}\right) - 2\log\xhz  - \left(1 - 2\xtz \right)
   \Phi\left(\frac{1}{4\xtz}\right)\bigg] ~, \\[16pt]
 \frac{d_e^{W\gamma}}{e} &=& \frac{2 e^2}{(16\pi^2)^2} \frac{y_e^\text{SM}}{\sqrt{2}v}
 ~\text{Im}\,\kappa_e \bigg[ \left( 1 + 6\xwh \right) \left( 2 +
   \log \xwh \right)  \nonumber \\ && \hspace{4cm} - \left(6 \xwh - 7 \right)
   \xwh \Phi\left(\frac{1}{4\xwh}\right) \bigg] \,, \\[16pt]
 \frac{d_e^{WZ}}{e} &=& \frac{e^2}{(16\pi^2)^2s_w^2} \frac{y_e^\text{SM}}{4\sqrt{2}v}
 ~\text{Im}\,\kappa_e \left( 1 - 4 s_w^2\right)  \left(1 - \xzh
 \right)^{-1} \nonumber \\ 
 && \times \bigg[ \left( 2 + 12\xwh - \frac{1}{c_w^2} - 2\xzh \right) \log\xzh
   \nonumber \\ && \qquad + \left(14 -12\xwh - \frac{3}{c_w^2} + 2\xzh
   \right) \xwh \Phi\left(\frac{1}{4\xwh}\right) \nonumber \\ 
 &&  \qquad + \left(2 + 12\xwh - \frac{1}{c_w^2} + \frac{4\xzh}{c_w^2} - 18\xzh
   \right) c_w^2 \Phi\left(\frac{1}{4c_w^2} \right) \bigg] ~.
\end{eqnarray}
\begin{eqnarray}
 \frac{d_e^{W}}{e} &=& \frac{e^2}{(16\pi^2)^2}
 \frac{y_e^\text{SM}}{18\sqrt{2} v}\frac{1}{s_w^2} ~ \text{Im}
 \kappa_e ~ \xhw \nonumber \\ 
  && \times \bigg\{6 \big(\xwh^2 + 4\xwh - 2\big)
 \Phi\left(\frac{1}{4\xwh}\right)  - 6 \big(4\xwh^3 + 3\xwh^2 - 4\big)
 \text{Li}_2(1-\xwh) \nonumber \\
  && \qquad - \pi^2 \xwh^2 \big(3 + 4\xwh \big) + 24\xwh
 \big(\xwh-1\big) + 24\xwh \big(\xwh+1\big) \log \xwh \nonumber \\ 
  && \qquad - 3 (4 \xwh^3 + 3\xwh^2 - 4)\log^2 \xwh \bigg\}~,
\end{eqnarray}
\begin{eqnarray}
 \frac{d_e^{Z}}{e} &=& \frac{e^2}{(16\pi^2)^2}
 \frac{y_e^\text{SM}}{36\sqrt{2}v} \frac{1}{s_w^2 c_w^2} ~\text{Im}
 \kappa_e~ \xhz (8c_w^4-12c_w^2+5) \nonumber \\ && 
  \times \bigg \{ - 6 \big( 4\xzh^3 + 3\xzh^2 - 1 \big) \text{Li}_2 (1-\xzh)
  - 3 \big( 1 - 2 \xzh - 8 \xzh^2 \big) \Phi\left(\frac{1}{4\xzh}\right) \nonumber \\ &&
  \qquad - \pi^2 \xzh^2 \big( 3 + 4\xzh \big) - 6 \xzh \big( 1 - 4\xzh
  \big) + 6 \xzh \big( 1 + 4\xzh \big) \log \xzh \nonumber \\ 
  && \qquad - 3 \big( 4\xzh^3 + 3\xzh^2 - 1 \big) \log^2 \xzh \bigg \}
  \nonumber \\ 
  && + \frac{e^2}{(16\pi^2)^2} \frac{y_e^\text{SM}}{6\sqrt{2}v} \frac{1}{c_w^2} ~\text{Im}
 \kappa_e~ (s_w^2-c_w^2) \xhz^3 \nonumber \\ && 
  \times \bigg \{12 \big( 1 - 4\xzh + \xzh^2 \big)
  \text{Li}_2 (1-\xzh) - 3 \big( 1 - 6 \xzh + 8 \xzh^2 \big)
  \Phi\left(\frac{1}{4\xzh}\right) \nonumber \\ && 
  \qquad - \pi^2 \big( 1 - 4\xzh \big) - 6 \xzh^2 + 12 \xzh^2 \log \xzh \nonumber \\
  && \qquad + 3 \big( 2\xzh^2 - 4\xzh +1 \big) \log^2 \xzh \bigg \}%
\end{eqnarray}
To simplify the expressions we defined the mass ratios $x_{ij} \equiv
M_i^2/M_j^2$, $c_w=M_W/M_Z$, and $s_w = \sqrt{1-c_w^2}$. 
The function $\Phi(z)$ is given by~\cite{Davydychev:1992mt}
\begin{equation}
\begin{split}
  \Phi(z) & = 4 \bigg( \frac{z}{1-z} \bigg)^{1/2} \text{Cl}_2 \big(2
  \arcsin(z^{1/2})\big) \,, \\
  \text{Cl}_2 (\theta) & = - \int_0^\theta dx \log |2 \sin (x/2)| \,,
\end{split}
\end{equation}
for $z<1$ and by
\begin{equation}
\begin{split}
  \Phi(z) & = \bigg( \frac{z}{z-1} \bigg)^{1/2} \bigg\{ -4 \text{Li}_2
  (\xi) + 2 \log^2 \xi - \log^2 (4z) + \frac{\pi^2}{3} \bigg\} \,, \\
  \xi & = \frac{1 - \big(\frac{z-1}{z}\big)^{1/2}}{2} \,,
\end{split}
\end{equation}
for $z>1$, where ${\rm Li}_2 (x) = -\int_0^x du \, \ln (1-u)/u$ is the
usual dilogarithm.
The numerical size of the individual contributions is
\begin{multline}
 \frac{d_e^\text{2loop}}{e}  = \frac{d_e^{t\gamma}}{e} +
 \frac{d_e^{tZ}}{e} + \frac{d_e^{W\gamma}}{e} + \bigg( \frac{d_e^{WZ}}{e} +
 \frac{d_e^{W}}{e} + \frac{d_e^{Z}}{e} \bigg) \\ 
 = \text{Im}\,\kappa_e \times \big( - 6.44 - 0.12 + 13.85 - 2.22 \big) 
 \times 10^{-27} \text{cm} \,.
\end{multline}
Note that the Barr-Zee contributions involving $Z$ bosons are
suppressed by the small vector coupling of the $Z$ boson to leptons
proportional to $(1 - 4 s_w^2)$.

We checked explicitly that the corresponding contributions to the
anomalous magnetic moment of the electron can be obtained via
\begin{equation}\label{eq:translate}
 \Delta a_e = \frac{(\text{Re}\,\kappa_e - 1)}{\text{Im}\,\kappa_e} 2
 m_e \left( \frac{d_e}{e} \right) \,.
\end{equation}
The results of this appendix can also easily be adapted to obtain
expressions for flavor violating dipole transitions such as $\mu \to e
\gamma$ in the presence of flavor-violating Higgs couplings.

%%%%%%%%%%%%%%%%%%%%%%%%%%%%%%%%%%%%%%%%%%%%%%%%%%%%%%%%%%%
\section{Enhanced Higgs production through a loop hole?} \label{sec:loophole}
%%%%%%%%%%%%%%%%%%%%%%%%%%%%%%%%%%%%%%%%%%%%%%%%%%%%%%%%%%%

%%%%%%%%%%%%%%%%%%%%%%%%%%%%%%%%%%%%%%%%%%%%%%%
\begin{figure}[t]
\begin{center}
\includegraphics[width=0.3\textwidth]{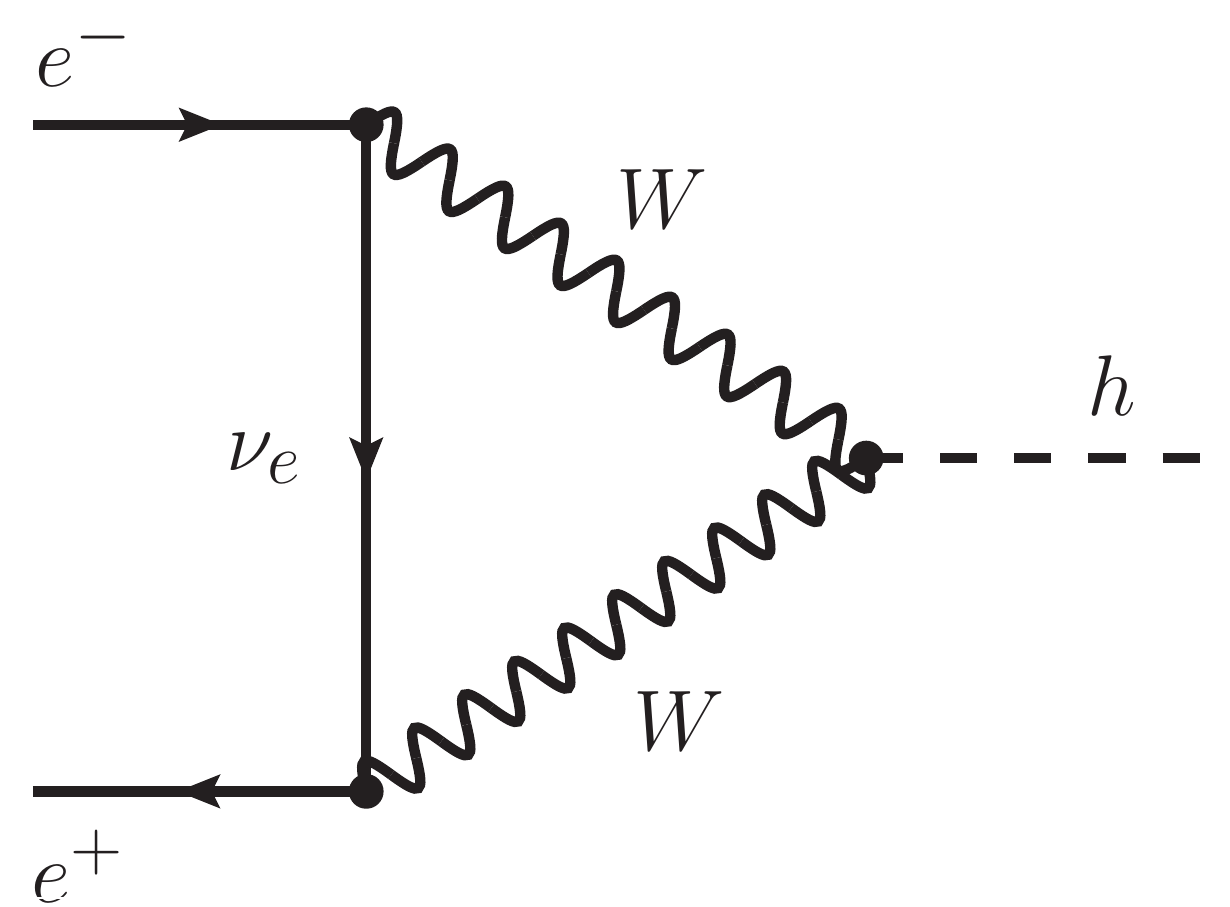}~~~~~~~~
\includegraphics[width=0.3\textwidth]{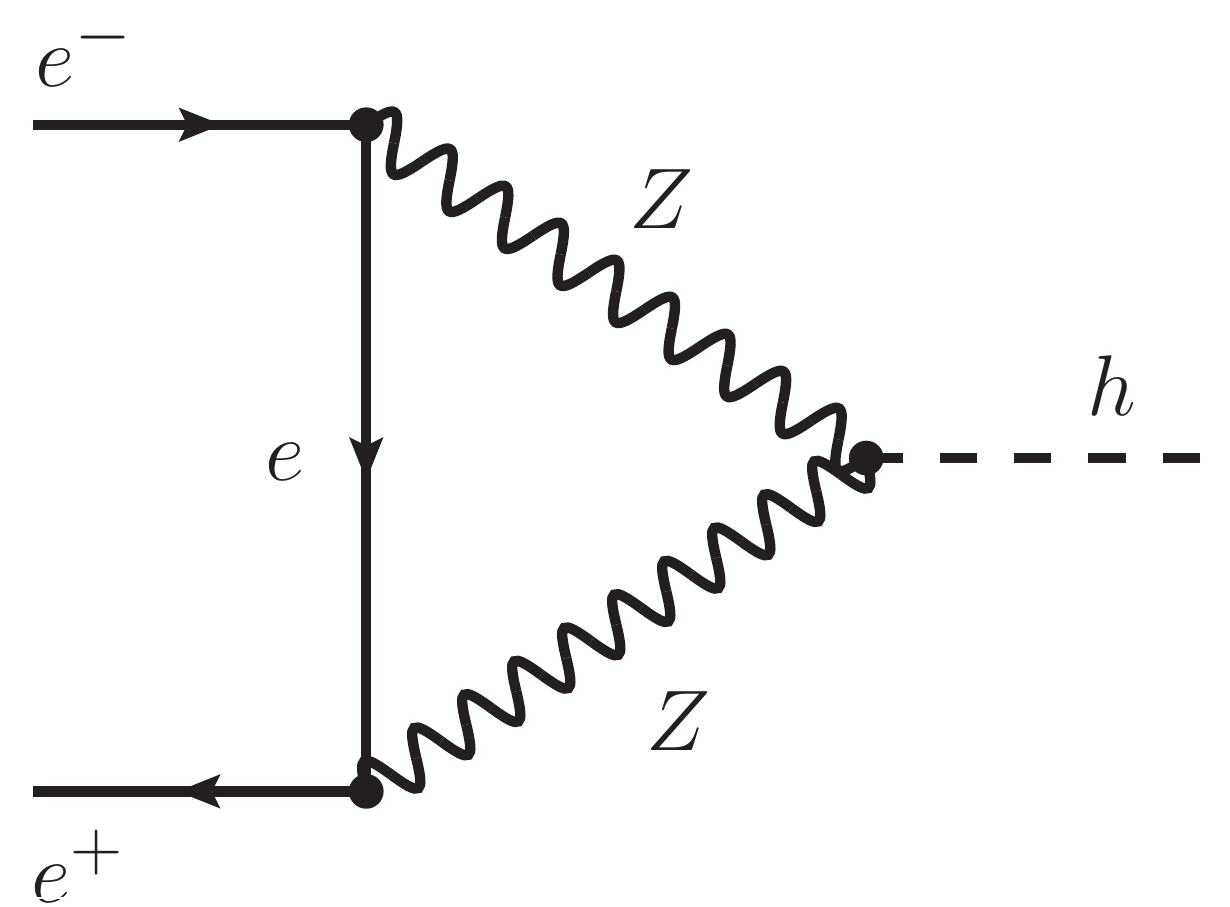}
\end{center}
\caption{Sample one-loop Feynman diagrams which naively look like they
might give a SM $s$-channel Higgs production cross section that is not suppressed by
  the small electron Yukawa. As shown in the text, chiral symmetry implies
that the amplitude for this process is suppressed by the electron mass for
on-shell external fermions. \label{fig:1loopeeh}}
\end{figure}
%%%%%%%%%%%%%%%%%%%%%%%%%%%%%%%%%%%%%%%%%%%%%%%

Naively one might expect that, in the SM, higher-order Feynman graphs
(for example Fig.~\ref{fig:1loopeeh}) could lead to enhanced
$s$-channel Higgs production, not suppressed by the small Yukawa
coupling. Here we show that this expectation is wrong. To see this
note that in the limit of vanishing electron Yukawa coupling (and
ignoring neutrino masses), the SM Lagrangian has an exact enhanced
(chiral) symmetry rotating the left- and right-handed components of
the electron field.  Thus any non-vanishing amplitude of electrons
coupling to the Higgs must either be proportional to the electron
Yukawa coupling (as we want to show) or preserve chiral symmetry.  But
any chiral symmetry preserving coupling of electrons has the electrons
combined into a vector which must be dotted into a electron momentum
to form a Lorentz invariant amplitude.  Then the electron equation of
motion can be used to turn the momentum into the electron mass.

To make this argument more explicit, consider the amplitude for the
transition of an on-shell electron-positron pair into a (not
necessarily on-shell) Higgs boson (the argument for the reverse
process is very similar). Its most general Lorentz spinor structure is
of the form
\begin{equation}
\bar v_e({\mathbf p}',\sigma') [ A + B \gamma_5 + C_\mu \gamma^\mu +
  D_\mu \gamma^\mu \gamma_5 + E_{\mu\nu} \sigma^{\mu\nu} ]
u_e({\mathbf p},\sigma) \,,
\end{equation}
where $A, B, C_\mu, D_\mu, E_{\mu\nu}$ are coefficients which must be
constructed out of the Lorentz invariants $p^2=p'^2=m_e^2$, $p\cdot
p'$ and the two independent Lorentz vectors $p_\mu$ and $p'_\mu$.

The amplitudes proportional to $A,B,E_{\mu\nu}$ violate chiral
symmetry, thus they can only be generated proportional to the electron
Yukawa coupling. The amplitudes with $C_\mu$ and $D_\mu$ preserve
chiral symmetry and need not be suppressed.  However, by Lorentz
symmetry $C_\mu$ and $D_\mu$ must be proportional to either $p_\mu$ or
$p'_\mu$. Thus we obtain amplitudes of the form $\bar v_e({\mathbf
  p'})\, \slashed{p} \,u_e({\mathbf p})$ and $\bar v_e({\mathbf p'})\,
\slashed{p'} \,u_e({\mathbf p})$ which are proportional to $m_e$ by
the equations of motion for on-shell external electrons. Therefore,
$s$-channel Higgs production in electron-positron collisions is always
suppressed by at least one power of $m_e$ (and possible loop factors).

This general argument based on spin and Lorentz invariance
continues to apply for amplitudes with additional {\it soft} photons
which cannot carry angular momentum.  However amplitudes with
additional hard photons need not be suppressed by the electron mass.
For example, the process $e^+e^- \rightarrow \gamma^* \rightarrow
\gamma\, h$ does arise in the SM and is not suppressed by the electron
mass. However it is suppressed by the small loop-induced coupling of the Higgs to
photons and was therefore not relevant for Higgs production at LEP2.
The presence of the hard photon in the final state would of course
allow experimenters to distinguish this process from the s-channel
Higgs production process in attempts to measure the Higgs-electron
coupling.

\end{appendix}

\newpage

%%%%%%%%%%%%%%%%%%%%%%%%%%%%%%%%%%%%%%%%%%%%%%%%%%%%%%%%%%%%%%%%%%%%%%%%%%

%\cleardoublepage
\phantomsection
\addcontentsline{toc}{section}{References}
\bibliography{paper}
  \bibliographystyle{kpinunu_Xt}

%%%%%%%%%%%%%%%%%%%%%%%%%%%%%%%%%%%%%%%%%%%%%%%%%%%%%%%%%%%%%%%%%%%%%%%%%%

\end{document}